# Direct measurement of the nonconservative force field generated by optical tweezers

Pinyu Wu,[1,*] Rongxin Huang,[1,*] Christian Tischer,[2] Alexandr Jonas,[3] and Ernst-Ludwig Florin[1,†]

[1] *Center for Nonlinear Dynamics and Department of Physics, University of Texas at Austin, Austin, Texas 78712, USA*
[2] *European Molecular Biology Laboratory, 69117 Heidelberg, Germany*
[3] *Institute of Scientific Instruments, Academy of Sciences of the Czech Republic, 61264 Brno, Czech Republic*

**Abstract**

The force field of optical tweezers is commonly assumed to be conservative, neglecting the complex action of the scattering force. Using a novel method that extracts local forces from trajectories of an optically trapped particle, we measure the three-dimensional force field experienced by a Rayleigh particle with 10 nm spatial resolution and femtonewton precision in force. We find that the force field is nonconservative with the nonconservative component increasing radially away from the optical axis, in agreement with the Gaussian beam model of the optical trap. Together with thermal position fluctuations of the trapped particle, the presence of the nonconservative force can cause a complex flux of energy into the optical trap depending on the experimental conditions.

---

[*] Both authors contributed equally
[†] email: florin@chaos.utexas.edu

*Introduction.*—Optical trapping has broad applications among researchers after nearly 40 years of development [1, 2]. Examples include simple manipulation of nanoparticles and cells [3-5], precise measurements of piconewton forces and nanometer or smaller displacements in biological systems [6-9], and three-dimensional imaging of polymer networks [10]. In optical tweezers, a laser beam is focused by a high numerical aperture objective lens to a diffraction-limited spot in which two types of forces act: a stabilizing gradient force that results from the intensity gradient and points towards higher intensity and a destabilizing scattering force that points along the propagation direction of the light (optical axis). Strong focusing generates a strong gradient force along the optical axis that eventually dominates over the scattering force and forms a stable three-dimensional trap. Neglecting the detailed action of the scattering force, single beam gradient traps (optical tweezers) have been commonly assumed over time to act as Hookean springs, creating a three-dimensional harmonic potential for the trapped particle [11]. Most force experiments in biology and physics have been performed under this assumption. Force experiments with optical tweezers are usually performed in two ways. In direct force measurements, the force is determined by the particle's displacement from its equilibrium position assuming a fixed spring constant for each direction. This type of experiment includes the measurement of the forces generated by the polymerase during transcription [12, 13] or by an individual motor protein [14-17] and the forces needed to unfold muscle proteins or RNA hairpins [18-20]. In indirect force measurements, the particle's spatial probability distribution is converted into an energy landscape using the Boltzmann distribution. Force-extension and stiffness-extension profiles are then calculated as first and second derivatives of the energy landscape. This method was initially used to calibrate optical tweezers [21] and was later also applied to investigate the mechanics of motor proteins in three dimensions [22]. An inherent assumption in indirect force experiments is that the trapped particle is in thermal equilibrium and explores the energy landscape only driven by thermal forces originating in the surrounding fluid. The thermal equilibrium or similar assumptions were made in experiments that studied the escape of a particle over an energy barrier [23], the violation of the second law of thermodynamics for small systems and short time scales [24], and the fluctuation theorems for nonequilibrium systems in statistical physics [25-27]. If the gradient force were the only force acting on a trapped particle, the thermal equilibrium assumption would be valid in all cases. However, the scattering force is always present in an optical trap along with the gradient force. In the Rayleigh regime, in which the particle diameter is much smaller than the wavelength of light, the scattering force is proportional to the local light intensity at the particle position and points along the propagation direction of the light. As the intensity drops strongly away from the optical axis, the scattering force drops too; this inhomogeneity of the scattering force subsequently generates a nonconservative contribution to the force field. Despite a recent attempt [28], the nonconservative force field in optical tweezers has never been measured directly due to a lack of experimental techniques.

In this letter, we first introduce a novel method for measuring a three-dimensional force field from the drift component of Brownian motion. We then determine experimentally the force field acting on a Rayleigh particle in single beam gradient trap and compare it with the theoretical result obtained using a Gaussian beam model for the light intensity distribution in the trap. Both, experiment and the Gaussian beam-based model show a significant nonconservative contribution to the force field that increases away from the optical axis.  Finally, we discuss effects of the nonconservative force on typical optical trapping experiments.

*Methods.*—To determine the force field of an optical trap experimentally, we need to introduce a method for measuring the local force acting on the trapped particle without assuming any particular property of the force field except that it is time-invariant. Typically, time series of the particle's thermal position fluctuations in the trap are readily available and, therefore, we want to calculate the force field directly from these thermal position fluctuations.  To achieve this, we consider the Brownian motion of a particle in an external force field within a fluid medium.  The equation of motion for such a particle with mass $m$ at position vector $\vec{r}$ is:

$$m\ddot{\vec{r}} = \vec{F}_{stoch} + \vec{F}_{fric} + \vec{F}_{trap} \qquad (1)$$

where $\vec{F}_{stoch}$ is the stochastic thermal force that drives the Brownian motion, $\vec{F}_{fric}$ is the viscous drag force in the fluid, and $\vec{F}_{trap}$ is the force generated by the optical trap.  For times much longer than the characteristic time scales of the particle's inertia and the hydrodynamic memory effect, the inertial term on the left hand side of (1) can be neglected. The viscous drag force then simplifies to Stokes' law, $\vec{F}_{fric} = -6\pi\eta a\vec{v}$, for a particle with a radius $a$ moving at velocity $\vec{v}$ in a fluid of viscosity $\eta$ [29].  The stochastic force drives the diffusion of the particle but does not change its average position. In contrast, the external trapping force $\vec{F}_{trap}$ leads to an average drift of the particle in its direction. Depending on the time scale of observation and the magnitude of the external force, the particle's motion is dominated either by the drift or by diffusion. Even for the motion in weak external fields where diffusion dominates, however, the random displacements average out by observing the particle for a sufficiently long time. Therefore, the external force field can be calculated as

$$\vec{F}_{trap}(\vec{r}_0) = 6\pi\eta a \frac{\langle \Delta \vec{r} \rangle_{\vec{r}=\vec{r}_0}}{\Delta t} \qquad (2)$$

where $\langle \Delta \vec{r} \rangle_{\vec{r}=\vec{r}_0}$, which we refer to as *local drift,* is the average displacement of the particle in a time interval $\Delta t$ when it starts at position $\vec{r}_0$ and moves under the external force $\vec{F}_{trap}(\vec{r}_0)$. In practical terms, the particle's average local drift can be calculated from a position time series in the following way: every time $t$ the particle visits a selected volume element at $\vec{r}_0$, the local drift at this volume element is calculated as the difference between its current position and its position at $t+\Delta t$, and the result is averaged over the total number of visits N to that particular volume element (see Fig. 1 for illustration)

$$\langle \Delta \vec{r} \rangle_{\vec{r}=\vec{r}_0} = \frac{\sum_{i=1}^{N} [\vec{r}_i(t+\Delta t) - \vec{r}_i(t)]}{N} \qquad (3)$$

Applying the local drift method to a Brownian particle in an optical trap is demanding for several reasons. First, in order to measure the correct magnitude of the local force that acts on the particle, the size of the volume element has to be small in comparison to the spatial variation of the force field. In our experiments we chose volume elements with an edge length of 10 nm. Additionally, in order to measure the displacement vectors $\vec{\Delta r}$ within such a small volume element, the position of the particle has to be measured with much higher precision than the volume element's dimensions. In Brownian motion, spatial precision and temporal resolution of the position measurement are directly coupled. For instance, a 200 nm particle in water at room temperature diffuses about 2 nm in 1.7 µs. Therefore, a sampling rate of at least 600 kHz is required to reach a 2 nm spatial precision. Only recently, such high precision and bandwidth has been achieved in three-dimensional particle tracking [29, 30]. The remaining task is to collect an adequate amount of position data for each volume element within the trapping volume. Fortunately, the optical trap confines the particle to a small volume and forces it to revisit the selected volume element over and over (Fig. 1). Hence, position data can be recorded repeatedly for the same volume element until sufficiently high precision for determining the average local drift and, consequently, the local force field is achieved.

*Experiment.*—In our experiments, we used a standard single beam gradient trap with a high bandwidth position detector as described before [29, 30]. A linearly polarized Nd:YAG laser beam (1064 nm, IRCL-850-1064-S, CrystaLaser) is focused by a high numerical aperture water immersion objective lens (UPlanSApo, NA=1.2, Olympus) into a fluid sample chamber to generate the optical trap. The laser power measured at the focal plane was 26 mW. A solution of 200 nm polystyrene beads with a concentration of approximately one bead in 100x100x100 µm$^3$ was prepared in deionized water. A single bead is optically trapped and its three-dimensional position is measured by forward scattered light interferometry [30]. The position signals in all dimensions are recorded simultaneously with a 16-bit data acquisition board (NI-6120, National Instruments) at a sampling rate of 600 kHz. For each trapped particle, 100 sets of position data are recorded with $10^6$ points per set. After the position data are calibrated [see supporting materials], the described local drift method is used to calculate the three-dimensional force field. We used up to $7\times10^5$ position measurements per volume element for calculating the experimental force field.

*Results.*—Fig. 2 (left) shows the projection of force field onto the transversal *x-y* plane at the average z position [31]. For small displacements from the optical axis, the magnitude of the force increases linearly with the displacement (see supplemental material). However, the force vectors generally do not point towards the optical axis because the force constants along the *x* and *y* axes differ by a factor of 1.7 ($k_x / k_y = 1.7$). The measured force constants $k_x = 6.2 \times 10^{-6}$ N/m and $k_y = 3.5 \times 10^{-6}$ N/m agree well with previous experimental data and theoretical calculations [32] in which the polarization of the trapping laser was taken into account. A weaker force constant is observed in the plane of polarization of the trapping laser (y plane) as expected. Fig. 2 (right) shows the projection of the force field onto the *x-z* plane at the average y position (y=0) [31]. The force along the optical axis increases much more slowly with displacement than that along the lateral directions and the ratios of the force constants ($k_x / k_z \sim 7$, $k_y / k_z \sim 4$) are also in good agreement with earlier measurements [32].

A standard way to quantify the local nonconservative component of a force field is to calculate its curl as $curl(\vec{F}) = \vec{\nabla} \times \vec{F}$, which is zero by definition for a conservative force field. Fig. 3a shows the projection of the three-dimensional curl of the experimental force field shown in Fig. 2 onto the x-y plane. Since we measured no significant axial component of curl, Fig. 3a represents the true magnitude and orientation of the curl field. The vortex-like structure of the curl field with counterclockwise orientation indicates that the experimental force field has indeed a significant nonconservative component. Because of its vector product nature, $curl(\vec{F})$ is perpendicularly oriented to the nonconservative force, which originates from the scattering force and points mainly along the z axis. Its counterclockwise orientation corresponds to a scattering force decreasing away from the optical axis. The curl field orientation would become clockwise if the scattering force increased away from the optical axis, as expected for instance for particles that are large relative to the wavelength of light. The center of the vortex appears at about 40 nm along the positive y-axis and the magnitude of the vectors increase away from this point. To verify our results, we performed first order Brownian dynamics simulations of a 200 nm particle moving in a single beam gradient trap formed by a Gaussian beam [see supplemental material]. The beam parameters were chosen to reflect quantitatively the experimental parameters. We calculated the force field from the simulated particle position tracks using the local drift method as applied to the experimental data before. As shown in Fig. 3b, the curl of the calculated force field agrees with high accuracy with the curl of the force field used in the Brownian dynamics simulation (Fig. 3b, inset), thus validating the precision of the local drift method. However, the advantage of comparing the experimental curl with the curl extracted from simulated position data (instead of the analytical curl field) is that statistical uncertainties are correctly reflected, i.e. less populated locations at the trap periphery will show a larger error. The curl of the experimental and simulated force field agrees in general well with a few exceptions. For the simulated data, the position of zero curl is located exactly on the optical axis, and there is no asymmetry between the x- and the y-axis. The shift of the position of vanishing curl in the experiment is very likely a result of the imperfect alignment of the optical trap. The asymmetry in the curl of the experimental force field is a result of the polarization dependence of the scattering force that was not taken into account in our simulation. Since the transversal beam intensity profile changes less steeply in the direction of polarization (y), a weaker change of the scattering force in this direction is expected and observed. Since the change of the scattering force with respect to the y-axis determines the magnitude of curl along the x-axis, a smaller curl component should be expected along the x-axis, which is clearly visible in Fig. 3a. From the curl we can estimate the magnitude of the effective nonconservative force locally [33]. For example, at a position 30 nm from the trapping center (x=0, y=-30 nm, z=0), the nonconservative force is approximately 2 fN.

To estimate the average work that can be done by the nonconservative force on the trapped particle, we integrate the force along different closed paths in the *x-z* plane (Fig. 4). For a particle following a rectangular closed path along the optical axis from z = -40nm to z = +40nm, and back on a path at x = 20 nm away from the optical axis (Fig. 4, blue path, right side), the energy put into the system is 0.25 $k_B T$. The average nonconservative force acting on the particle along this path can be estimated from $<F_{nc}>=W/s$, where $W$ is the work done along the path and $s$ is the length of the closed path. For the discussed case, the average nonconservative force is 5 fN which corresponds to an average particle speed of 3 μm/s. With this considerable speed, the particle would circle the path about 15 times a second and put approximately 3.6 $k_B T$ per second into the system.

However, we would like to point out that a particle is unlikely to follow such a path spontaneously. If no thermal forces acted on the trapped particle, the dominating gradient force would just pull it back to the point of zero force, regardless of its starting position in the trap. The energy would be dissipated by the viscous force and the particle would come to rest. No circulating motion of the particle would be observed in this case, unlike the circulation one would expect for a particle in a vortex. Therefore, energy due to the action of the nonconservative force can only be transferred to a particle continuously through the action of thermal forces that drive the particle away from the position of zero force.

In order to discuss a situation where the nonconservative force may play an important role, we consider again the curl of the experimental force field (Fig. 3a). The magnitude of the curl increases away from its minimum position. Therefore, the strongest effect is expected in experiments where the trapped particle is displaced far away from the optical axis. This is the case, for instance, in single molecule force experiments when large forces (10-100 pN) are applied to pre-stretch or unfold molecules [7, 12, 18]. However, much larger particles are typically used in these experiments to achieve high forces, when the Rayleigh regime approximation is no longer valid. In fact, a geometrical optical force calculation shows that the curl field pattern for large particles can even reverse, meaning that the scattering force increases with the distance from the optical axis, and therefore, the curl of its force field is expected to change its orientation. More precise calculations are required for the transition regime where the particle diameter is on the order of the wavelength of the trapping laser, but the experimental verification of the nonconservative effect of the scattering force in this regime is straightforward with the method we have described.

*Summary.*—In summary, we have developed a novel method to precisely measure the three-dimensional force field of an optical trap from the trajectories of the Brownian motion of a trapped particle with nanometer spatial resolution. Our method imposes no requirements about the nature of the probed force field as long as it is constant over the course of the experiment. We confirmed that the force field generated for a Rayleigh particle in a single beam gradient trap is nonconservative as predicted by the Gaussian beam model. In combination with thermal position fluctuations of the particle, the nonconservative forces lead to a complex flow of energy into the system. The actual flow depends on the particular experiment and requires a theoretical case-by-case analysis.

Quantifying the drift component of Brownian motion presents a novel way to measure weak forces on the nanometer scale that were previously obscured by thermal fluctuations. Because of the coupling of temporal and spatial resolution in the observation of Brownian motion, these experiments require position detectors with both high bandwidth and spatial precision. With recent progress in detector technology for optical tweezers [34], sub-nanometer precision in mapping three-dimensional force fields might be within reach and would pave the way for a new class of experiments in single molecule biophysics and the study of Brownian motion in confined geometries.

**Acknowledgement:**

This work was supported by the National Science Foundation grants No. DBI 0052094 and PHY 0647144.

**Figure Captions:**

**Fig. 1.** A local force acting on a particle in an optical trap can be determined from the time series of its Brownian motion. (a) A single beam optical trap formed by a focused Gaussian beam propagating in the positive z direction. The Gaussian profile (red) represents the intensity distribution of the beam. The particle diffuses in the trap and crosses a volume element (blue) multiple times (not drawn to scale). (b) Time series of the particle position along the x-axis. The two horizontal lines indicate the position of the boundaries of the volume element shown in (a) along the x-axis. Note: Only the sections of the position signal labeled in blue and green actually cross the selected volume element. (c) Multiple paths crossing the same volume element (left) are analyzed to obtain the local force from the average drift component of the diffusing particle (right).

**Fig. 2.** Experimental force field calculated from the local drift of the trapped particle. Left: force field in the transversal x-y plane located at the average z position (trapping center). Right: force field in the x-z plane located at the average y position. Please note that the transversal force vectors usually do not point towards the center of the trap as a result of the polarization dependence of the trapping stiffness. Experimental parameters: 200 nm diameter polystyrene particle trapped at a laser power of 26 mW at the focal plane. The laser was polarized along the y-axis. The force field was calculated using Eq. 2 and 3 with a time interval $\Delta t=17$ μs.

**Fig. 3.** Nonconservativeness of the experimental, simulated and theoretical force field for a Rayleigh particle in a single beam gradient trap (optical tweezers). (a) Curl of the experimental force field calculated from the data shown in Fig. 2. This curl field is an average over a ±90 nm range along the optical axis around the average z position. The red circle indicates the approximate minimum position of the curl field. (b) Curl of the force field calculated from simulated position data. The Brownian dynamics simulation was performed with a 200 nm particle with n=1.57 at a laser power of 25 mW. Inset: curl of the theoretical force field applied in the simulation.

**Fig. 4.** Work done on a trapped particle by the nonconservative force along different paths. The experimental force field shown in Fig. 2b was used. The work for each path is given in units of $k_BT$.

**Figure 1:**

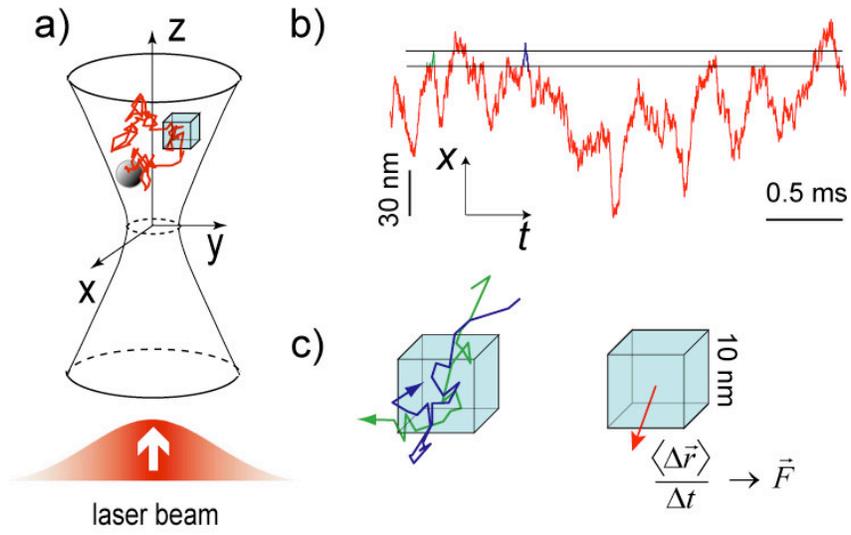

**Figure 2:**

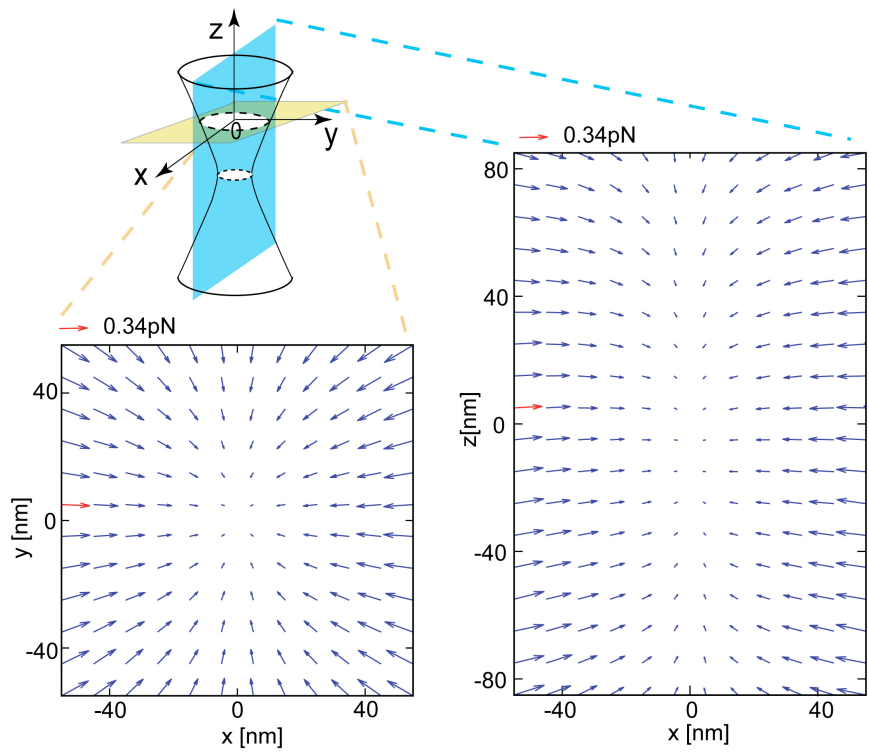

**Figure 3:**

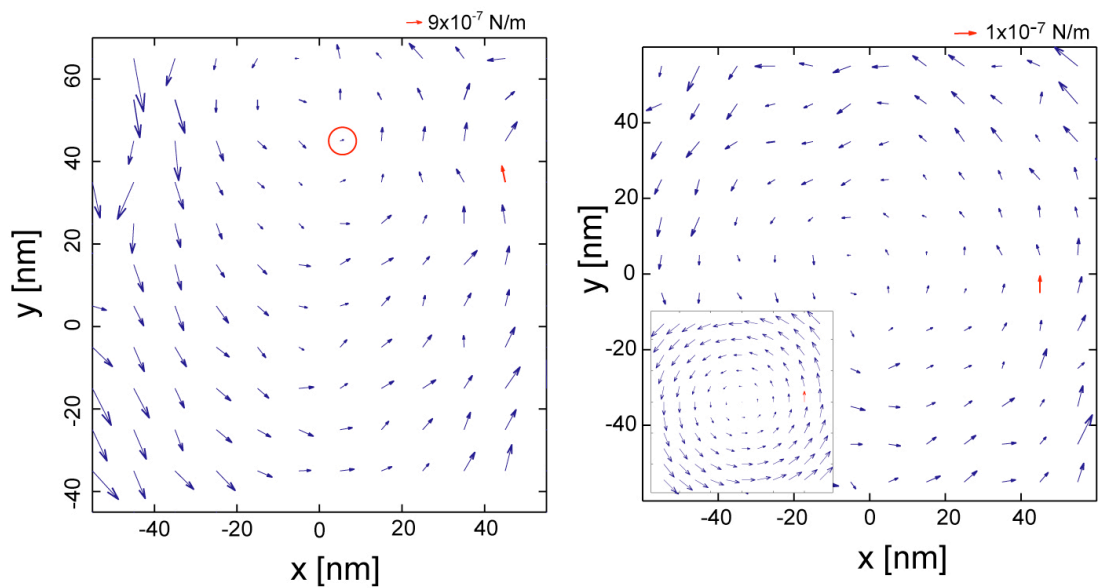

**Figure 4:**

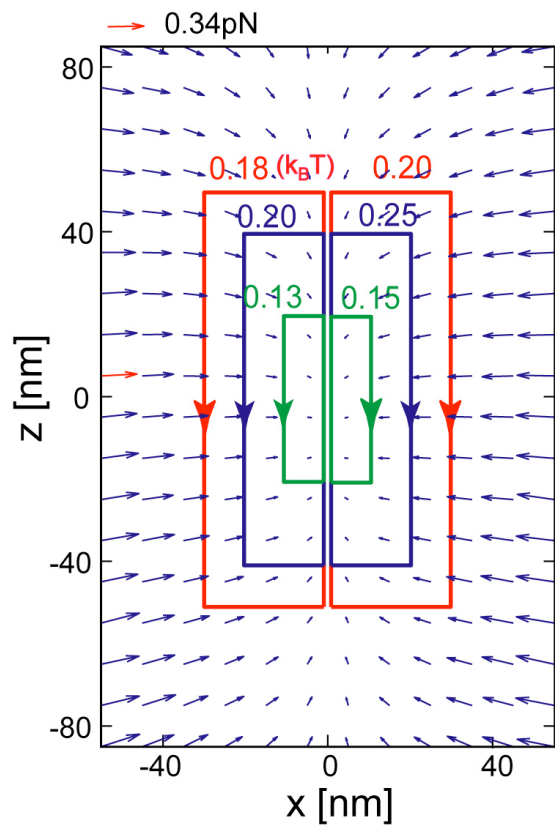